\documentclass[12pt]{article}
\usepackage{graphicx}
\usepackage{amsmath,amssymb}
\usepackage{graphics,color,array,calc,rotating,epsfig,psfrag}
\numberwithin{equation}{section}
\usepackage{cite}
\usepackage{bm}
\usepackage{dcolumn}  
\usepackage{amsfonts}
\usepackage[mathscr]{eucal}
\def\be{\begin{equation}} \def\ee{\end{equation}}
\def\bea{\begin{eqnarray}} \def\eea{\end{eqnarray}}
\def\bs{\boldsymbol}
\def\lb{\label} 
\def\a{\alpha}
 \def\b{\beta}

\newcommand{\beqar}{\begin{eqnarray*}}
\newcommand{\eeqar}{\end{eqnarray*}}

\newcommand{\prd}{\partial}

\begin{document}
\baselineskip 18pt%
\begin{titlepage}
\vspace*{1mm}%
\hfill%
\vspace*{15mm}%
\hfill
\vbox{
    \halign{#\hfil         \cr
          } 
      }  
\vspace*{20mm}

\centerline{{\Large {\bf Black Holes in Born-Infeld Extended New Massive Gravity }}}
\centerline{{\Large{\bf    }}}
\vspace*{5mm}
\begin{center}
{ Ahmad Ghodsi\footnote{ahmad@ipm.ir, a-ghodsi@um.ac.ir} and Davood Mahdavian Yekta\footnote{da.\,mahdavianyekta@stu-mail.um.ac.ir}}\\
\vspace*{0.2cm}
{ Department of Physics, Ferdowsi University of Mashhad, \\
P.O. Box 1436, Mashhad, Iran}\\
\vspace*{0.1cm}
\end{center}

\begin{abstract} 
In this paper we find different types of black holes for the Born-Infeld extended new massive gravity. Our solutions include (un)charged warped $(A)dS$  black holes for four and six derivative  expanded action. We also look at the black holes in unexpanded BI action. In each case we calculate the entropy, angular momentum and mass of the black holes. We also find the central charges for the CFT duals.
\end{abstract} 

\end{titlepage}

\section{Introduction}
The black hole solutions of three dimensional gravity have been investigated during recent years. The first black hole solutions, known as the BTZ black holes, were found in \cite{Banados:1992wn}. These solutions were found in the presence of a negative cosmological constant. Adding the higher derivative terms to the Einstein-Hilbert action in the presence of a cosmological constant changes the solutions and their asymptotic behaviors and their physical properties.

The topological massive gravity (TMG) describes propagation of the massive gravitons around the flat, de Sitter or anti de Sitter background metrics. This theory is constructed by adding a parity-violating Chern-Simons term to the Einstein-Hilbert action \cite{Deser:1982vy}. The cosmological TMG solutions contain either the BTZ black holes \cite{Banados:1992wn} or the warped $AdS_3$ black holes \cite{Moussa:2003fc}. The charged black hole solutions for topologically massive (TMGE) presented in \cite{Moussa:2008sj}. 

The new massive gravity (NMG) was found in \cite{Bergshoeff:2009hq}. This theory was constructed by adding a parity-preserving higher derivative term to the tree level action. In the NMG theory there are  the BTZ and  warped $AdS_3$ solution too, \cite{Clement:2009gq}. 
The charged black hole solutions for new massive gravity (NMGE) is given in \cite{Ghodsi:2010gk}.

Several attempts have been done to extend the three dimensional gravity theories to the higher curvature corrections. One of the most recent extensions is the Born-Infeld extension of the new massive gravity \cite{Gullu:2010pc}
\bea\label{BINMG}
S&=&\frac{2m^2}{\kappa}\int d^3 x\Big(\sqrt{-det(g_{\mu\nu}+\frac{\sigma}{m^2}\,G_{\mu\nu}+ a F_{\mu\nu})}-(1+\frac{\Lambda}{2m^2})\sqrt{-det g_{\mu\nu}}\Big)
\cr &&\cr
&+&\frac{\mu}{2}\int d^3 x\epsilon^{\mu\nu\rho}\!A_{\mu}\partial_{\nu}A_{\rho},
\eea
where we have added a gauge field strength and a Chern-Simons term in order to study the charged solutions of this theory\footnote{One may consider other terms here. For example consider a more general form as $aF_{\mu\nu}\rightarrow aF_{\mu\nu}+b F^2 g_{\mu\nu}+c {F_{\mu}}^\rho F_{\rho\nu}$. Our computations shows a similar behavior with a complicated form of solutions depending on $a, b$ and $c$ parameters.}.
In the above action $g_{\mu\nu}$ is the metric and $G_{\mu\nu}={\cal R}_{\mu\nu}-\frac12 {\cal R} g_{\mu\nu}$ is the Einstein tensor for the curvature tensor ${\cal R}$. $F$ is the field strength for a $U(1)$ gauge field. We consider $m$ as a mass parameter and $a$ and $\mu$ as two constant parameters, here $\kappa=8\pi G_N$. To have a positive coefficient for the scalar curvature we choose $\sigma=-1$.

By inserting $a=\mu=0$, and to second order of expansion for the small curvature parameter of the above action, one finds the new massive gravity action in \cite{Bergshoeff:2009hq}. The next order terms add other extensions to the new massive gravity, which are consistent with deformations of NMG obtained from AdS/CFT correspondence \cite{Sinha:2010ai}. The other extension of the NMG using AdS/CFT method in 3D is given by \cite{Paulos:2010ke}. The uncharged $AdS$ black hole solutions for this theory has been found in \cite{Nam:2010dd} (see also \cite{Alishahiha:2010iq}).

In this paper and in section two, we review a method for finding the black hole solutions for this theory. This method has been used in \cite{Clement:2009gq} to find the Warped solutions in new massive gravity. In sections three and four we expand the action (\ref{BINMG}) to four and six derivative terms. By solving the equations of motion we find the charged and uncharged black holes. We discuss the domain of validity of each solution and find the physical parameters (mass, angular momentum, temperature and entropy) of our black holes. We show that at each order of expansion the behavior of the solutions (the physical parameters) depend on different range of the parameters of the theory. In section five we do the same steps of sections three and four but for the BI action (\ref{BINMG}) (unexpanded action). We show that the behavior of the solutions is totally different when one considers the BI action (so we need to study the theory at each order of expansion to find the behavior of the black holes at that order). In section six we find the central charges for the CFT duals.
In section seven we summarize our results.
\section{How to find a black hole solution?}
To find the stationary circularly symmetric solutions for the Lagrangian (\ref{BINMG}) at different orders we use the dimensional reduction procedure, presented in \cite{Moussa:2008sj}, \cite{Clement:2009gq}. We use the following ansatz for the metric and the $U(1)$ gauge field
\be  \label{metr}
ds^2=\lambda_{ab}(\rho)\,dx^a dx^b + \zeta^{-2}(\rho)R^{-2}(\rho) \,d\rho^2\,,
\qquad {\cal A} = A_a(\rho) \,dx^a\,, 
\ee
where ($a,b=0,1$) and ($x^0 = t$, $x^1 = \varphi$). The parameter $\lambda$ can be expressed as a $2 \times 2$ matrix
\be
\lambda = \left(
\begin{array}{cc}
T+X & Y \\
Y & T-X
\end{array}
\right),
\ee
and $R^2\equiv \bs X^2=-T^2+X^2+Y^2$ is the Minkowski pseudo-norm of the ``vector'' $\bs X(\rho) = (T,\,X,\,Y)$.
We do the following steps to find the solutions:

1. We insert the ansatz (\ref{metr}) into the Lagrangian and reduce the action to $I = \int d^2x \int d\rho \, L$. By variation with respect to $A_a, \zeta, T, X $ and $Y$ we find the equations of motion.

2. The black hole solutions for equations of motion  in step 1 can be found by choosing the following ansatz for the vector field ${\bs X}$ and the gauge field ${\cal A}$ (this behavior is general, for example see the gauge field solutions in \cite{Moussa:2008sj} and \cite{Ghodsi:2010gk})
\be  
{\bs X}={\bs \alpha}\rho^2+{\bs \beta}\rho+{\bs \gamma}\,,\quad {\cal A}=c\,\Big(2z dt-(\rho+2\omega z)\, d\varphi\Big)\,. \label{X}
\ee
Inserting this ansatz into the equations of motion for $T, X$ or $Y$ or putting into the equation of motion for $\zeta$, we always find two conditions ${\bs \alpha}^2={\bs \alpha} \cdot {\bs \beta}=0$. We will find the values of $c$ and $z$ for the above ansatz by using the equations of motion.

3. Without lose of generality we choose $\zeta(\rho)=1$ in all  equations of motion.

4. Following to \cite{Clement:2009gq} we can choose a rotating frame and a length-time scale such that 
\be
{\bs \alpha}=(\frac12,-\frac12,0)\,,\quad
{\bs \beta}=(\omega,-\omega,-1)\,,\quad {\bs \gamma}=(z+u,z-u,-2\omega z)\,,
\ee
where $u=\frac{ \b_0^2 \rho_0^2}{4z}+\omega^2z$. By these parameters
one finds $R^2=(1-2z)\rho^2+{\bs \gamma}^2 \equiv  \b_0^2(\rho^2- \rho_0^2)$.
By the above parameters, we are able to write the metric in the warped-$AdS_3$ ADM form \cite{Clement:2009gq}
\be\lb{admmetr} 
ds^2 = - \b_0^2\frac{\rho^2- \rho_0^2}{r^2}\,dt^2 + r^2\Big[d\phi
  - \frac{\rho+(1- \b_0^2)\omega}{r^2}\,dt\Big]^2 + \frac1{ \b_0^2\zeta^2}\frac{d\rho^2}{\rho^2- \rho_0^2}\,,
\ee 
where $\rho_0$ (describes the location of the horizon) together with $\omega$ are two parameters of the theory and $r^2 = \rho^2 +2\omega\rho +\omega^2\,(1- \b_0^2) +\frac{ \b_0^2 \rho_0^2}{1- \b_0^2}$.

5. As indicated in \cite{Clement:2009gq}, in order to avoid the closed time-like solutions, we must have $0<\beta_0^2<1$ or $0<z<\frac12$. We impose this condition in order to find the domain of validity of our solutions. For charged solutions we consider the reality condition for the gauge field strength.

6. For each black hole solution we can find the entropy according to the Wald's formula 
\bea 
S &=& 4\pi A_h\Big(\frac{\delta{\cal L}}
{\delta {\cal R}_{0202}}(g^{00}g^{22})^{-1}\Big)_h\,,
\eea
where $A_h =
\frac{2\pi}{\sqrt{1-\b_0^2}}[\rho_0 + \omega(1-\b_0^2)]$ is the area of the horizon.

7. The computation of the mass and angular momentum is possible if we could linearize the field equations and use the ADT approach \cite{Abbott:1981ff} and \cite{Deser:2002rt}. Equivalently, we can follow the Cl\'{e}ment's approach in \cite{Clement:2009gq}. The main idea of this approach is the fact that the Lagrangian has $SL(2,R)$ symmetry. This symmetry allows us to write the metric as (\ref{metr}).
By this symmetry we can find a conserved current and its conserved  charge. Under infinitesimal symmetry transformation we can find the transformations of the gravitational and electromagnetic fields. By using these transformations we can find the conserved current, called the super angular momentum vector, ${\bs J}$. We have discussed the details of computing the super angular momentum in the appendix A. 
 
The conserved charge is the angular momentum and it can be read from the super angular momentum by using the relation $J=2\pi(\delta \bs J^T-\delta \bs J^X)$.
Here $\delta \bs J$ is the difference between the values of the super-angular momentum for the black hole and for the background solution.
The background solution is given by the values $ \rho_0=\omega=c=0$. Using these values we can find from (\ref{admmetr}) a horizon-less background metric 
\be 
ds^2=- \b_0^2dt^2+\rho^2 \big(d\phi-\frac{1}{\rho}dt\big)^2+\frac{1}{ \b_0^2 \zeta^2}\frac{d\rho^2}{\rho^2}\,,
\ee
with $
 \bar{{\bs \a}}=(\frac12,-\frac12,0)\,, \bar{{\bs \b}}=(0,0,-1)\,, \bar{{\bs \gamma}}=(z,z,0)$.
To compute the mass, one could use the first law of the thermodynamics for black holes in the modified Smarr-like formula \cite{Moussa:2008sj}, which is appropriate for the warped $AdS_3$ black holes, i.e,
\be \label{fl}
M=T_HS+2\Omega_hJ\,.
\ee
Using the ADM form of the metric we can read the Hawking temperature and the horizon angular velocity (see \cite{Moussa:2008sj})
\be\label{TH}
T_H = \frac{\zeta\b_0^2 \rho_0}{A_h}\,,\quad \Omega_h =
\frac{2\pi\sqrt{1-\b_0^2}}{A_h}\,.
\ee
Inserting (\ref{TH}) into (\ref{fl}) and by using the value of the angular momentum we can find the mass of the black holes.
For every solution, we must check that, the physical parameters satisfy the differential form of the first law of the thermodynamics for black holes, i.e, 
\be
dM=T_H dS+\Omega_h dJ\,.
\ee
In our solutions, the physical parameters of mass, entropy and angular momentum are functions of two free parameters, $\rho_0$ and $\omega$, so the differentiations are with respect to these two free parameters.
For more details of computations see the appendix B.

{\bf{Note:}} For every Lagrangian we will use, we find three types of black holes. The first one is the uncharged black hole when $c=\mu=0$. The second one is the Maxwell charged (M-charged) black hole, where we consider $a\neq0$ but $\mu=0$. The third one is the Maxwell-Chern-Simons charged  black hole (MCS-charged). In the later case we consider $a\neq0$ and $\mu\neq0$, but to write the Lagrangian in the canonical form we choose $a^2=-\frac{\kappa}{2m^2}$  and $\mu=1$. Note that we are able to perform all the calculations for general values of $a$ and $\mu$.
\section{The four derivative action}
Expanding the general Lagrangian (\ref{BINMG}) up to four derivative terms, gives us the following Lagrangian ($Tr(AB)=A_{\mu\nu}B^{\nu\mu}$)
\bea
L_4
\!\!\!&=&\!\!\!\frac{2m^2}{\kappa}\sqrt{-g}\Big[-\frac{\sigma}{4m^2}R-\frac{\sigma^2}{4m^4}\Big(Tr(R^2)-\frac38R^2\Big)+\frac{\sigma a^2}{2m^2}\Big(Tr(RF^2)-\frac38RTr(F^2)\Big)\nonumber\\
\!\!\!&-&\!\!\!\frac14 a^2Tr(F^2)-\frac18 a^4\Big(Tr(F^4)-\frac14(Tr(F^2))^2\Big)-\frac{\Lambda}{2m^2}\Big]+\frac{\mu}{2}\epsilon^{\mu\nu\rho}A_{\mu}\partial_{\nu}A_{\rho}\,.
\eea
The equations of motion coming form the variation with respect to the gauge field are 
\be
\frac{\delta L_4}{\delta A_t}=\frac{2c}{\kappa}(a^4c^2m^2z+(z+m^2+\frac18)a^2+\frac12\mu\kappa)=0\,,\quad \frac{\delta L_4}{\delta A_{\varphi}}=\frac{\delta L_4}{\delta A_{\rho}}=0\,.
\ee
Variation with respect to $\zeta$ gives
\be
\frac{\delta L_4}{\delta \zeta}=\frac {1}{\kappa m^2}\{-a^2 c^2z(2+3a^2c^2z)m^4+(\frac14+\Lambda-\frac74a^2c^2z-4a^2c^2z^2)m^2
+\frac{1}{64}-\frac32z-z^2\Big\}\!=\!0,
\ee
and finally variations with respect to $T ,X$ and $Y$ give the following equations 
\bea
\frac{\delta L_4}{\delta T}\!\!\!\!&=&\!\!\!\!\frac{\delta L_4}{\delta X}\!=\!\frac{-1}{16\kappa m^2}\Big(17+8z-8m^2+2a^2c^2m^2(8m^2+9+12z)+16a^4c^4m^4z\Big)\!=\!0\,,\cr &&\cr
\frac{\delta L_4}{\delta Y}\!\!\!\!&=&\!\!\!\!0\,.
\eea
\subsection*{Uncharged solution:}
In this case we find the known NMG solution \cite{Clement:2009gq}
\be
z = -\frac{17}{8}+m^2\,,\quad \Lambda=\frac{1}{16}\frac{21-48m^2+16m^4}{m^2}\,.
\ee
With the above values, the domain of validity of the solution for $m^2$ will be
$
\frac{17}{8}< m^2 < \frac{21}{8}$ and for the cosmological constant it is $ 1.806<\Lambda<4.081\,.
$

The super angular momentum is given by
\bea
{\bs J}_{BH}\!\!\!\!&=&\!\!\!\!-\frac{(8m^2-21)(8m^2-17)}{32\kappa m^2}(1+\omega^2,1-\omega^2,2\omega)+\frac{\rho_0^2}{2\kappa m^2}\frac{8m^2-21}{8m^2-17}(1,-1,0)\,.\nonumber\\ 
\eea
Using the method reviewed in section 2, we can find the entropy, the angular momentum and the mass as follows
\bea\label{unch}
S&=&\frac{A_h}{2m^2 G_N}\,,\qquad
J=\frac{8m^2-21}{4 G_N m^2}(\frac{\rho_0^2}{{8m^2-17}}-\frac{{8m^2-17}}{16}\omega^2)\,,\cr &&\cr 
M&=&-\frac{(8m^2-21)(8m^2-17)}{32G_N m^2}\omega\,.
\eea
\subsection*{M-charged solution:}
In this case we consider $a\neq0$ and $\mu=0$. The equations of motion give the following solution
\be
z = -1-\frac{1}{8m^2}\,,\quad \Lambda=\frac{1}{16}\frac{-3-24m^2+16m^4}{m^2}\,,\quad c^2a^2={1-\frac{1}{m^2}}\,.
\ee
The domain of validity of this solution for $m^2$ will be
$-\frac{1}{8}< m^2 < -\frac{1}{12}$, which looks totally different from the uncharged case. The cosmological constant is limited between $-0.125<\Lambda<0.667$, so it has both plus and minus signs. It changes its sign at $m^2=\frac34-\frac{\sqrt{3}}{2}\cong-0.116$. We also find from the above solution that $9< c^2a^2< 13$.
Using the method in section 2 the super angular momentum can be found as
\bea
{\bs J}_{BH}\!\!\!\!&=&\!\!\!\!-\frac{(8m^2+1)(12m^2+1)}{32\kappa m^4}(1+\omega^2,1-\omega^2,2\omega)+\frac{\rho_0^2}{2\kappa}\frac{12m^2+1}{8m^2+1}(1,-1,0),
\eea
and the entropy, angular momentum and  mass are given by
\bea
S&=&\frac{A_h}{2 G_N}\,,\qquad
J=\frac{12m^2+1}{4G_N}({\frac{\rho_0^{2}}{8m^2+1}}-{\frac {8m^2+1}{16m^4}}\omega^2)\,,\cr &&\cr 
M&=&-\frac{(12m^2+1)(8m^2+1)}{32m^4G_N}\omega.
\eea
\subsection*{MCS-charged solution}
In this case we consider $a^2=-\frac{\kappa}{2m^2}$ and $\mu=1$. We find the following parameters from the equations of motion
\be
z =-\frac{1}{12}(1+\frac{1}{m^2})\,,\quad \Lambda=-\frac{1}{48}(16+\frac{1}{m^2})\,,\quad \kappa c^2=\frac{2-m^2}{1+m^2}\,.
\ee
The domain of validity of this solution will be
$-1< m^2 < -\frac{1}{7}$. Inserting this into the cosmological constant relation, we find that it is always negative $-0.312\cong-\frac{5}{16}< \Lambda <-\frac{3}{16}\cong -0.187$ in this domain. Finally we find from the solution that $\frac52< \kappa c^2$.

The super angular momentum for MCS-charged black hole is given by 
\bea
{\bs J}_{BH}\!\!\!&=&\!\!\!-\frac{(7m^2+1)(m^2+1)}{144\kappa m^4}(1+\omega^2,1-\omega^2,2\omega)+\frac{\rho_0^2}{4\kappa}\frac{7m^2+1}{m^2+1}(1,-1,0),
\eea
and the entropy, angular momentum and  mass are as follows
\bea
S&=&\frac{A_h}{4 G_N}\,,\qquad
J=\frac{7m^2+1}{8G_N}({\frac{\rho_0^{2}}{m^2+1}}-{\frac {m^2+1}{36m^4}}\omega^2)\,,\cr &&\cr
M&=&-\frac{(7m^2+1)(m^2+1)}{144m^4G_N}\omega.
\eea

As we see, each of the above solutions has its own domain of validity for $m^2$ and the cosmological constant. In all the above cases the entropy is proportional to the area of the horizon. The angular momentum and mass have the same functionality in terms of $\omega$ and $\rho_0$ but with different coefficients.
\section{The six derivative action}
By expanding the Lagrangian (\ref{BINMG}) up to six derivative terms we find the the following Lagrangian
\bea
L_6&=&L_4+\frac{2m^2}{\kappa}\sqrt{-g}\Big[\frac{\sigma^3}{6m^6}\Big(Tr(R^3)-\frac98RTr(R^2)+\frac{17}{64}R^3\Big)
\nonumber\\
&-&\frac{\sigma^2a^2}{m^4}\Big(\frac34Tr(R^2F^2)-\frac58RTr(RF^2)+\frac{19}{128}R^2Tr(F^2)-\frac{1}{16}Tr(R^2)Tr(F^2)\Big)
\nonumber\\
&+&\frac{\sigma a^4}{2m^2}\Big(Tr(RF^4)-\frac{7}{16}RTr(F^4)+\frac{7}{64}R(Tr(F^2))^2-\frac14 Tr(RF^2)Tr(F^2)\Big)
\nonumber\\
&-&\frac{a^6}{12}\Big(Tr(F^6)-\frac38 Tr(F^2)Tr(F^4)+\frac{1}{32}(Tr(F^2))^3\Big)\Big]\,.
\eea
The gauge field equations of motion are given by
\bea
\frac{\delta L_6}{\delta A_t}\!\!&=&\!\!\frac{3c}{\kappa m^2}\Big[a^6c^4m^4z^2+\frac23a^4c^2m^2z(z+\frac58+m^2)+\Big(\frac23m^4+(\frac23z+\frac{1}{12})m^2\nonumber\\
\!\!&-&\!\!\frac13z^2+\frac14z+\frac{1}{64}\Big)a^2+\frac13m^2\kappa\mu\Big]=0\,,\quad\frac{\delta L_6}{\delta A_{\varphi}}=\frac{\delta L_6}{\delta A_{\rho}}=0.
\eea
Variation with respect to $\zeta$ gives
\bea
\frac{\delta L_6}{\delta\zeta}&=&{\frac{2}{\kappa m^4}}\bigg[
{z^{3}+{\frac{9z^{2}}{16}}-\frac{3z}{32}}+{\frac{1}{1024}}
+\Big(\frac{3a^{2}c^{2}}{2}z^{3}-(a^{2}c^{2}+\frac12)z^2-\frac34({\frac{13a^{2}c^{2}}{32}}+1)z\nonumber\\
&+&{\frac{1}{128}}\Big)m^{2}-\Big(2a^{4}c^{4}(z+{\frac{29}{32}})z^{2}+2a^{2}c^{2}(z+{\frac{7}{16}})z\!-\!{\frac{1}{8}}-\frac{\Lambda}{2}\Big)m^{4}
-\Big(\frac{5a^{6}c^{6}}{2}z^{3}\nonumber\\
&+&\frac{3a^{4}c^{4}}{2} z^{2}+a^{2}c^{2}z\Big)m^{6}
\bigg]=0\,,
\eea
and variations of the Lagrangian with respect to $T ,X$ and $Y$ give the following relations
\bea
\frac{\delta L_6}{\delta T}&=&\frac{\delta L_6}{\delta X}=
\frac{2}{\kappa m^4}\bigg(\frac{3}{8}(z^{2}+{\frac{5}{4}}z-{\frac{11}{64}})-\frac12\Big(\frac32a^{4}c^{4}z^{2}+a^{2}c^{2}z+1\Big)a^{2}c^{2}m^{6}
\nonumber\\
&-&\Big(\frac{5}{8}a^{4}c^{4}(z+{\frac{13}{10}})z+\frac34a^{2}c^{2}(z+\frac34)-{\frac{1}{4}}\Big)m^{4}
\nonumber\\
&+&\Big(\frac{1}{2}(z^{2}-{\frac{9}{16}}z-{\frac{51}{128}})a^{2}{{c}}^{2}-{\frac{17}{32}}-\frac{1}{4}z\Big)m^{2}
\bigg)=0\,,\nonumber\\
\frac{\delta L_6}{\delta Y}&=&0.
\eea
We now try to solve the above equations of motion and find different (un)charged black hole solutions. As the previous section we will find the super angular momentum, and by computing the value of the entropy from the Wald's formula and by using the first law of the thermodynamics for black holes we will be able to find the angular momentum and the mass of our solutions. Our calculations for all the solutions show that, there is a general behavior. In fact we find that the super angular momentum can be written as
\bea\label{genjbh}
{\bs J}_{BH}=\frac{1}{2\kappa m^4}U(1+\omega^2,1-\omega^2,2\omega)+\frac{1}{4\kappa m^4z}V\rho_0^2(1,-1,0)\,,
\eea
and the entropy, angular momentum and  mass are given by
\be\label{genSJM}
S=\frac{A_h}{4 G_N m^4}\frac{V}{z-\frac12}\,,
J=\frac{1}{4G_Nm^4}(U\omega^2+\frac{1}{2z}V\rho_0^2)\,,
M={\frac{\omega z(U\omega
-V\rho_0)}{{G_N}m^{4}({\rho_0}+2\omega z)}}\,,
\ee
where $U, V, z$ and $A_h$ depend on the solution.
\subsection*{Uncharged solution:}
For uncharged black holes, when $a=0$ one finds the following values for $z$ and $\Lambda$ \footnote{This special case has been found in \cite{Nam:2010dd}. Here we find the domain of validity of their solution and the domain of the cosmological constant.}
\bea\label{z}
m^2&=&\frac{1311}{1448}+\frac{3}{181} z\pm\frac{9}{724}(11695-208z-320z^2)^\frac12\,\\
\label{lambdasol} \!\!\!\!\Lambda&=&\frac{1}{m^4}\Big(\frac{45}{32}+\frac54m^2-\frac73m^4+\frac{16}{27}m^6\pm\frac{5}{54}(m^4-\frac{27}{16})\Big(81+144m^2-80m^4\Big)^{\frac12}\Big)\,.\nonumber\\
\eea
To have a causally regular warped black hole one needs to consider $0<z<\frac12$, so we find the following domains by looking at equation (\ref{z}):

{\bf{The upper sign of (\ref{z}):}} In this case 
$2.247< m^2<2.250$. The lower bound happens at $z=\frac12$ and the upper bound is located at the point of $z=\frac18$. 
Inserting this domain into the equation (\ref{lambdasol}), for the upper sign of (\ref{lambdasol}) we find
$-0.176<{\Lambda}<-0.167$
and for the lower sign of (\ref{lambdasol}) one obtains
$-0.167<{\Lambda}<-0.034$.

{\bf{The lower sign of (\ref{z}):}} In this case
$-0.420<m^2<-0.043$.
Here the lower bound happens at $z=\frac12$ and the upper bound is located at $z=0$.
Inserting again the above domain into the equation (\ref{lambdasol}), for the upper sign of (\ref{lambdasol}) one finds
$1.348<{\Lambda}<1.500$
and for the lower sign we have
$2.368<{\Lambda}<3.325$.

The super angular momentum (\ref{genjbh}) and the entropy, angular momentum and mass in (\ref{genSJM}) all are given by the following values for $U$ and $V$
\bea \label{UV1}
U&=&z(m^4-(\frac18+5z)m^2-\frac{1}{128}-\frac58z+\frac{11}{2}z^2)\,,\nonumber\\
V&=&(z-\frac12)(m^4-(\frac18+z)m^2-\frac{1}{128}-\frac18 z+\frac32 z^2)\,.
\eea
\subsection*{M-charged solution:}
To find the Maxwell charged black holes we consider $a\neq0$. We find the following equation for the value of $z$ in terms of $m$
\bea\label{poly1}
\!\!\!& &\!\!\!6144(z+1)^{2}m^{8}\!-\!\Big(8192z^{3}+15360z^{2}+5376z-1792\Big)m^{6}+\Big(8192z^{4}+14336z^{3}\nonumber\\
\!\!\!&+&\!\!\!5376z^{2}-640z+272\Big)m^{4}+\Big(1024z^{3}+768z^{2}-16z+24\Big)m^{2}-15z^{2}-12z=0\,.\nonumber\\
\eea
By finding the roots of the above polynomial one is able to find the values for $\Lambda$ and $c$. For simplicity we define $\Delta=\Big(-20m^{4}+2(1-8z)m^{2}+16z^{2}-4z+1\Big)^\frac12$ then
\bea\label{Lambda3}
\Lambda\!\!\!&=&\!\!\!\frac{1}{1728m^4}\bigg[1024m^{6}-2016m^{4}+\Big(1536z^{2}+768z-336\Big)m^{2}-2048z^{3}-384z^{2}\nonumber\\
\!\!\!&-&\!\!\!96z-5\mp32\Big(10m^{4}-4(z+1)m^{2}+16z^{2}+5z+1\Big)\Delta\bigg]\,,\\
\label{c3}
c^2a^2\!\!\!&=&\!\!\!\frac{1}{24zm^2}\Big(-8m^2-8z-5\pm 4\Delta\Big)\,,
\eea
where the plus sign in (\ref{Lambda3}) corresponds to minus sign in (\ref{c3}) and vice versa.

To find the domain of validity for this black hole one needs to find the roots of (\ref{poly1}). A numerical analysis shows that there are two real solutions for $m^2$. The figure 1 (left) shows the results of this numerical analysis. 
For the negative roots we find that $-0.167<m^2<-0.100$. For the positive roots we have $0<m^2<0.038$, where in this case the extremum point is near to the point $z\cong0.180$.

The numerical calculations show the following values of the cosmological constant (see the figure 2).

{\bf{The upper sign:}}
For the equation  (\ref{Lambda3}) with the upper sign the cosmological constant is always negative (left diagram in figure 2). 
For $m^2<0$ we have $-54.821<\Lambda<-0.635$
and for $m^2>0$ we find $\Lambda<-38.026$, where the extremum is located at $z\cong 0.124$.

{\bf{The lower sign:}}
The numerical analysis shows that for the equation  (\ref{Lambda3}) with the lower sign the cosmological constant is always positive (the right diagram of the figure 2). 
For $m^2<0$ we have $0.229<\Lambda<0.826$ and for $m^2>0$ we have $1.763<\Lambda$, where the extremum is located at $z\cong 0.347$.

It remains to find the behavior of $c^2a^2$ in equation (\ref{c3}).
For the upper or lower signs of (\ref{c3}) one could find either positive or negative values (see the figure 3).
Since the value of $c$ must be real ($c^2>0$) so depending on $a^2$ sign one may choose either the left or the right diagram.

The superangular momentum (\ref{genjbh}) and the values for the entropy, angular momentum and  mass in equation (\ref{genSJM}) are given by the following values 
\bea\label{UV2}
U\!\!\!&=&\!\!\!\frac13 z\Big((2z-m^2)\Delta
+14m^4+2(1-4z)m^2+8z^2-z+\frac{3}{16}\Big)\,,\nonumber\\
V\!\!\!&=&\!\!\!\frac19(z-\frac12)\Big((2z-m^2-\frac14)\Delta+14m^4+(1-8z)m^2+8z^2-2z+\frac{5}{16}\Big).
\eea
\subsection*{MCS-charged solution:}
The Maxwell-Chern-Simons black holes with $a^2=-\frac{\kappa}{2m^2}$ and $\mu=1$ are given by the following relation
\bea\label{poly2}
& &z(5z+4)-\frac{1024}{3}\Big(z^{3}+\frac34z^{2}-\frac{1}{64}z+\frac{3}{128}\Big)m^{2}-4096\Big(z^{4}+\frac56z^{3}-\frac{13}{384}z^{2}\nonumber\\
&-&\frac{77}{1536}z+\frac{7}{384}\Big)m^{4}+4096\Big(z^{3}+\frac{23}{24}{z}^{2}-\frac{25}{192}z-\frac{11}{192}\Big)m^{6}-1792\Big(z^{2}+\frac{59}{42}z
\nonumber\\
&+&\frac{73}{336}\Big)m^{8}-3072\Big(z+\frac{1}{12}\Big)m^{10}=0,
\eea
by finding the roots of the above polynomial again, one could find the values for $\Lambda$ and $c$. We define $\Delta=\Big(4m^4+2(1-8z)m^2+16z^2-4z+1\Big)^\frac12$, then we find
\bea\label{Lambda4}
\Lambda&=&\
\frac{1}{1728m^4}\Big[-1280m^{6}-(1440+1152z)m^{4}+\Big(1536z^{2}+768z-336\Big)m^{2}\cr &&\cr
&-&2048z^{3}-384z^{2}-96z-5\mp32\Big(20m^{4}+4(z+1)m^{2}-16z^{2}-5z-1\Big)\Delta\Big]\,,\cr &&\cr  &&
\eea
and
\bea
\label{c4}
\kappa c^2&=&\frac{2}{3z}(m^2+z+\frac58\pm \frac12\Delta)\,,
\eea
where the plus sign in (\ref{Lambda4}) corresponds to minus sign in (\ref{c4}) and vice versa.

Similar to previous sections we can find the domain of validity of our solution. Looking at the equation (\ref{poly2}) one finds that this equation has two negative and one positive roots, figure 1 (right). 

Inserting these roots into the relation (\ref{Lambda4}) one finds: 

{\bf{The positive root:}}
For $m^2>0$ we have $\Lambda>0$ for upper sign of (\ref{Lambda4}) and for the lower sign  we find $\Lambda<0$ (see the figure 4). 

{\bf{The negative roots:}}
For $m^2<0$ we have both signs (depending on $z$). For upper sign of (\ref{Lambda4}) we can draw the figure 5 (left) and for lower sign we find the figure 5 (right).

The $U$ and $V$ values in this case are given by
\bea \label{UV3}
U\!\!\!\!&=&\!\!\!\!\frac13z\Big((2m^4+m^2-2z)\Delta
+4m^6+4(z+\frac{15}{8})m^4+2(1-4z)m^2+8z^2-z+\frac{3}{16}\Big)\,,\nonumber\\
V\!\!\!\!&=&\!\!\!\!\frac19(z-\frac12)\Big(-(2z-m^2-\frac14)\Delta+11m^4+(1-8z)m^2+8z^2-2z+\frac{5}{16}\Big)\,.
\eea

\section{All order solution}
As we saw in previous sections, at each level of expansion we have different properties for our solutions. It is interesting to consider the BI action  (\ref{BINMG}) without expansion and find its physical properties too.
If we use our ansatz then we will find the following sets of equations of motion:
  
The gauge field equations of motion are given by
\be
\frac{\delta L}{\delta A_t}=\frac{c}{\kappa}\Big(\mu\kappa+4a^2m^3\Big(\frac{-1+4m^2+8z}{1-8m^2+16(1-2a^2c^2z)m^4}\Big)^{\frac12}\Big)=0\,,\quad
\frac{\delta L}{\delta A_{\varphi}}=\frac{\delta L}{\delta A_{\rho}}=0.
\ee
Variation with respect to $\zeta$ gives
\bea
\frac{\delta L}{\delta\zeta}&=&\frac{2m^2}{\kappa}(1+\frac{\Lambda}{2m^2})\cr &&\cr
&-&\frac{16m^6-8(1-(2+a^2c^2)z-4a^2c^2z^2)m^4+(1+8z)m^2-3z}{\kappa m\Big((4m^2+8z-1)(16(1-2a^2c^2z)m^4-8m^2+1)\Big)^\frac12}=0\,,
\eea
and variations of the Lagrangian with respect to $T ,X$ and $Y$ give the following relations
\bea
\frac{\delta L}{\delta T}&=&\frac{\delta L}{\delta X}=\frac{-8a^2c^2m^6+4(1-6(z+\frac14)a^2c^2)m^4-10m^2+\frac94}{\kappa m\Big((4m^2+8z-1)(16(1-2a^2c^2z)m^4-8m^2+1)\Big)^\frac12}
=0\,,\nonumber\\
\frac{\delta L_4}{\delta Y}&=&0\,.
\eea
\subsection*{Uncharged solution}
Inserting $a=\mu=0$ in the above equations of motion we find (see also \cite{Nam:2010dd})
\bea\label{all1}
&&\frac{-4m^4+(1-4z)m^2-3z+m(2m^2+\Lambda)(4m^2+8z-1)^\frac12}{\kappa m(4m^2+8z-1)^\frac12}=0\,,\nonumber\\
&&\frac{4m^2-9}{4\kappa m (4m^2+8z-1)^\frac12}=0\,.
\eea
The solution for these equations is
\be
m=\pm\frac32\,,\qquad \Lambda=\frac{12+8z-9\sqrt{2(z+1)}}{2\sqrt{2(z+1)}}\,.
\ee
Unlike the previous cases here the value of $m$ is fixed.
To have a regular black hole we need either $
-\frac92+3\sqrt{2}<\Lambda<0$ when $0< z< -\frac{15}{64}+\frac{9}{64}\sqrt{17}$ or $0<\Lambda<-\frac92+\frac83\sqrt{3}$ when $-\frac{15}{64}+\frac{9}{64}\sqrt{17}< z< \frac12$.

The entropy of such a black hole can be found by using the Wald formula for $m=\pm\frac32$
\be \label{unchBI}
S
=\pm\frac{A_h}{3G_N\sqrt{2(z+1)}}\,.
\ee
As we see in order to have a positive value entropy we must choose $m=\frac32$.

We can find the super angular momentum as before
\be
{\bs J}
=\pm\frac{2z}{3\kappa}\frac{2z-1}{\sqrt{2(z+1)}}\Big((1+\omega^2,1-\omega^2,2\omega)
-\frac{1}{4z^2}\rho_0^2(1,-1,0)\Big)\,.
\ee
Using the above results the angular momentum and the mass of the solution can be found as
\be
J
=\pm\frac{(2z-1)}{12G_N z\sqrt{2(z+1)}}\Big(\rho_0^2-4z^2\omega^2\Big)\,,\quad M=\mp\frac{2\omega z(2z-1)}{3G_N\sqrt{2(z+1)}}\,.
\ee
\subsection*{M-Charged solution}
Solving the equations of motion when $a\neq0$ the only consistent solution to the equations of motion will be 
$m^2=\frac14$. Inserting this value to other equations of motion gives $z=0$ and $c=0$ with $\Lambda=-\frac12$. So in this case the gauge field vanishes and we have not charged black hole.
\subsection*{MCS-Charged solution}
Solving equations of motion gives the following solution
\be
z=-\frac{1}{48}\frac{4m^2+3}{m^2}\,,\quad \Lambda=-\frac13\,,\quad \kappa c^2=\frac{9-4m^2}{3+4m^2}\,. 
\ee
As we see in this case the value of cosmological constant is fixed.
The values of $m^2$ are limited between $-\frac34<m^2<-\frac{3}{28}$, also we find $\frac{11}{3}<\kappa c^2$.

The super angular momentum for this black hole is equal to
\be
{\bs J}
=-\frac{1}{2304}\frac{(3+28m^2)(3+4m^2)}{\kappa m^4}(1+\omega^2,1-\omega^2,2\omega)
+\frac{3+28m^2}{4\kappa(3+4m^2)}\rho_0^2(1,-1,0)\,.
\ee
The entropy, angular momentum and  mass of this black hole are given by
\be
S=\frac{A_h}{4G_N},
J=\frac{3+28m^2}{8G_N}\Big({\frac{{{\rho_0}}^{2}}
{3+4m^2}}-{\frac{(3+4m^{2}){\omega}^2}{576m^4}}\Big), M=-\frac{(3+4m^2)(3+28m^2)}{2304G_N m^4}\omega.
\ee
\section{A note on the central charges}
In previous sections we found the warped $AdS_3$ solutions. It is natural to ask what are the properties of the 2 dimensional CFT duals to these solutions. In the pure Einstein gravity, the global $SO(2,2)$ symmetry is enhanced to two
copies of an infinite dimensional Virasoro algebra with  the $SL(2,R)_L\times SL(2,R)_R$ symmetry which has the central
charge \cite{Brown:1986nw}, $c_L=c_R=\frac{3\,l}{2\,G_3}\,, $ 
where $l$ is the length of $AdS$ space.
When one considers the higher derivative terms or add the gauge fields, the value of the central charge changes. Some properties of the CFT dual theories for warped $AdS_3$ and BTZ black holes have been found in \cite{Nam:2010dd},\cite{Alishahiha:2010iq},\cite{Garbarz:2008qn},\cite{Fareghbal:2010yd}. The isometery group of the dual CFT in the asymptotic limit of the warped AdS black hole is $SL(2,R)_R \times U(1)_L$,\cite{Anninos:2008fx}. To find the central charge we use the Cardy's formula. 
\be \label{card1} 
S_{BH}=\frac{\pi^2}{3}\big(\,c_L\,T_L+c_R\,T_R\,\big)\,,
\ee
where $T_{L/R}$ are the left and right temperatures which we define them for this kind of solutions as
\be \label{lrtemps} 
T_L\equiv\frac{1-2z}{2\pi\sqrt{2z}}\rho_0\,,\quad\quad T_R\equiv\frac{1-2z}{2\pi\sqrt{2z}}2\omega z\,.
\ee 
One can also find the same results for central charge by using another form of the Cardy's formula $S_{BH}=2\pi\,\big(\sqrt{\frac{c_L}{6}\,E_L}+\sqrt{\frac{c_R}{6}\,E_R}\,\big)$ where $E_L$ and $E_R$ are the left and right energies defined in terms of physical mass and angular momentum of the black hole, i.e. $E_L\equiv\frac12 M \omega-J$ , $E_R\equiv\frac12 M \omega$ \cite{Ghodsi:2011ua}.

For each type of solution we can find separate central charges. For the uncharged solution by substituting (\ref{lrtemps}) into (\ref{card1}) and using the entropies found in (\ref{unch}) and (\ref{unchBI}) we obtain
\be c^{(4)}=\frac{48}{G(1-2z)(8z+17)}\,,\quad c^{(BI)}=\frac{4}{G(1-2z)\sqrt{2(z+1)}}\,,\ee
similarly for the M-charged solution we have
\be c_M^{(4)}=\frac{6}{G(1-2z)}\,,\quad c_M^{(BI)}=\frac{4}{G(1-2z)\sqrt{2(z+1)}}\,,\ee
and finally for the MCS-charged solution they become
\be c_{MCS}^{(4)}=\frac{3}{G(1-2z)}\,,\quad c_{MCS}^{(BI)}=\frac{3}{G(1-2z)}\,.\ee
For the sixth derivative Lagrangian in terms of parameter $V$, we find the following central charge
\be c^{(6)}=-\frac{6V}{(1-2z)^2\,m^4}\,\ee
where $V$ are given  by (\ref{UV1}),(\ref{UV2}) and (\ref{UV3}). Note that in this paper we have considered the case where $\mu=1$, but in general one can keep it in equations and find the central charges in terms of $\mu$, \cite{Ghodsi:2011ua}.  
\section{Summary and Discussion}
In this paper we have found different black hole solutions for the Born-Infeld extension of New Massive Gravity. We have extended the NMG in two directions, gravity and electromagnetism. The electromagnetic part contains Maxwell term and Chern-Simons term.
We have found three types of warped $(A)dS$  solutions. These are Uncharged, Maxwell charged and Maxwell-Chern-Simons charged black holes. For each of them we have found the domain of validity (CTC-free and reality of the gauge field strength). We have found these black holes for expanded (up to four and six derivative) and unexpanded BI action. The physical properties of the solutions in each case are totally different and one cannot find the four or six derivative properties form the BI properties.

In this theory we have two types of parameters. The first type includes parameters in the Lagrangian. The $m^2$ which is the mass parameter for our massive gravity theory and the cosmological constant $\Lambda$. The second type includes the parameters which are coming from the solutions. From (\ref{admmetr}) we see three parameters. The parameter $\beta_0$ or equivalently $z$ which is limited by CTC free condition, and two free parameters $\rho_0$ and $\omega$. There is another parameter $c$, in (\ref{X}), which corresponds to the electric or magnetic charge of the solutions.

Among these parameters in all the solutions, the values of $m^2$, $\Lambda$ and $c$ are controlled by the value of the $z$ parameter.
Our computation shows that each of these parameters may have positive or negative values. So we may have de Sitte or anti de Sitte solutions in our theory which depends on the level of expansion and existence of the M-charges or MCS-charges in the theory.
Our results is summarized as follows:

In four derivative case we have only one set of roots for $m^2$ which is either positive or negative depending on the solution. But in six derivative case for uncharged and M-charged cases we have two sets of roots one positive and one negative. For MCS-charged solutions we have three sets of roots, one positive and two negative.
In BI case the situation is totally different. For uncharged case we find only a single point for $m^2$. For the BI case there is not M-charged solution and for MCS-charged black holes there is only one set of roots for $m^2$.

In four derivative action for uncharged solutions $m^2>0$ and $\Lambda>0$. For M-charged solutions $m^2<0$ but $\Lambda$ changes its sign in a specific point $z$. For MCS-charged black holes $m^2<0$ and $\Lambda<0$.

In six derivative action for uncharged solutions $m^2$ and $\Lambda$ can have both signs, this is true for M-charged solutions. For MCS-charged black holes we have the same behavior but there are situations where $\Lambda$ changes its sign in a specific value of $z$.

In unexpanded BI action, for uncharged solutions $m^2$ has fixed value and the cosmological constant has both plus and minus signs. For M-charged solutions there is no solution. For MCS-charged black holes $m^2<0$ and the cosmological constant has a fixed value $\Lambda=-\frac13$.

For all the above solutions we have found the entropy, angular momentum and mass. Our results satisfy the differential form of the first law of the thermodynamics for black holes. We have used the free parameters of the theory, $\rho_0$ and $\omega$ to show this.

In all of the solutions, the entropy is proportional to the area of the horizon. Using the Cardy's formula we have found the central charges of the CFT duals. For our solutions in despite of presence of Maxwell-Chern-Simons term the left and right central charges are equal. In general as shown in \cite{Ghodsi:2011ua} if there is a gravitational Chern-Simons term (TMG model) then this equality will be broken and we obtain different left and right central charges.
\appendix
\section{The super angular momentum}

To compute the angular momentum it is essential to note that the Lagrangian has $SL(2,R)$ symmetry and the super angular momentum is a $SL(2,R)$ conserved current. Under the infinitesimal transformation, we find the following transformation for the fields
\bea
&&\Delta T= \epsilon^1 Y-\epsilon^2 X\,,\quad
\Delta X= \epsilon^0 Y-\epsilon^2 T\,,\quad
\Delta Y=-\epsilon^0 X+\epsilon^1 T\,,\nonumber\\
&&\Delta A_0= \frac12(\epsilon^0+\epsilon^1) A_1-\frac12\epsilon^2 A_0\,,\quad
\Delta A_1= \frac12(-\epsilon^0+\epsilon^1) A_0+\frac12\epsilon^2 A_1\,,\quad \Delta A_2=0\,.\nonumber\\
\eea
Using the above relations one can find the conserved currents (Super angular momentum). For the gravity part we find
\bea
\bs{J}_{Gr}=\Big[\!\!\!\!&+&\!\!\!\!(\frac{\partial L}{\partial X'}Y-\frac{\partial L}{\partial Y'}X)
-((\frac{\partial L}{\partial X''})'\,Y-(\frac{\partial L}{\partial Y''})'\,X)
+(\frac{\partial L}{\partial X''}Y'-\frac{\partial L}{\partial Y''}X'),\nonumber\\
&+&\!\!\!\!(\frac{\partial L}{\partial T'}Y\,+\frac{\partial L}{\partial Y'}T)
\,-((\frac{\partial L}{\partial T''})'\,Y\,+\,(\frac{\partial L}{\partial Y''})'\,T)
\,+(\frac{\partial L}{\partial T''}Y'\,+\frac{\partial L}{\partial Y''}T'),\nonumber\\
&-&\!\!\!\!(\frac{\partial L}{\partial T'}X+\frac{\partial L}{\partial X'}T)
+((\frac{\partial L}{\partial T''})'\,X\,+\,(\frac{\partial L}{\partial X''})'\,T)
-(\frac{\partial L}{\partial T''}X'+\frac{\partial L}{\partial X''}T')
\Big]\,,\nonumber\\
\eea
where prims denote the derivatives with respect to $\rho$ variable. One may write this result as $\frac{\delta L}{\delta\bs{X}}\wedge {\bs X}$ (see \cite{Clement:2009gq}). Also for the electromagnetic part one finds
\bea
\bs{J}_{EM}=\frac12\Big[(\frac{\partial L}{\partial A_0'}A_1-\frac{\partial L}{\partial A_1'}A_0)\,,
(\frac{\partial L}{\partial A_0'}A_1+\frac{\partial L}{\partial A_1'}A_0)\,,
-(\frac{\partial L}{\partial A_0'}A_0-\frac{\partial L}{\partial A_1'}A_1)
\Big]\,.
\eea
The total super angular momentum then will be $\bs{J}=\bs{J}_{Gr}+\bs{J}_{EM}$.
\section{The differential form of the first law}
As we mentioned in section 2 the black hole solutions in this paper have two free parameters $\rho_0$ and $\omega$.
In order to check that the entropy, angular momentum and mass satisfy the differential form of the first law of the thermodynamics for black holes we need to know every parameter in terms of these two free parameters. 

The area of the horizon is equal to $ A_H=\frac{2\pi}{\sqrt{2z}} (\rho_0+2\omega z)$.  The temperature of the black holes is given by $T_H=\frac{(1-2z)}{2\pi}\frac{\sqrt{2z} \rho_0}{\rho_0+2\omega z}$, and the angular velocity can be read from the metric and it is equal to $\Omega_H=\frac{2z}{\rho_0+2\omega z}$.
For all solutions in this paper we must check the following relations 
\bea\label{dform}
\frac{\prd M}{\prd \rho_0}- T_H \frac{\prd S_{BH}}{\prd \rho_0}- \Omega_H \frac{\prd J}{\prd \rho_0}=\,0\,,\qquad
\frac{\prd M}{\prd \omega}- T_H \frac{\prd S_{BH}}{\prd \omega}- \Omega_H \frac{\prd J}{\prd \omega}=\,0\,.
\eea
As an example (we have checked the above relations for every solution, the computations are very lengthy in most cases) for uncharged solutions in four derivative case we find
\bea
&&\frac{\prd  M}{\prd \rho_0}- T_H \frac{\prd S_{BH}}{\prd \rho_0}- \Omega_H \frac{\prd J}{\prd \rho_0}=-\frac12\frac{\rho_0 (8m^2-17-8z)}{(\rho_0+2\omega z)G m^2 (8m^2-17)}=\,0\,\,,\nonumber\\
&&\frac{\prd M}{\prd \omega}- T_H \frac{\prd S_{BH}}{\prd \omega}- \Omega_H \frac{\prd J}{\prd \omega}=-\frac14 \frac{ (z-\frac{21}{8}+m^2)(8m^2-17-8z)}{(\rho_0+2\omega z) G m^2}=\,0,
\eea
where we have used the equation of motion $z=m^2-\frac{17}{8}$ in this case. As another example for uncharged solution in the six derivative case we find 
\bea
&&\frac{\prd M}{\prd \rho_0}- T_H \frac{\prd S_{BH}}{\prd \rho_0}- \Omega_H \frac{\prd J}{\prd \rho_0}=\nonumber\\
&&-\frac{1}{64} \frac{\omega^2 z^3 (124m^4-128m^2 z-272m^2+192z^2+240z-33)}{(\rho_0+2\omega z)^2 G m^4}=\,0\,\,\,,\nonumber\\
&&\frac{\prd M}{\prd \omega}- T_H \frac{\prd S_{BH}}{\prd \omega}- \Omega_H \frac{\prd J}{\prd \omega}=\nonumber\\
&&\frac{1}{64} \frac{\omega z^3 \,\rho_0 (124m^4-128m^2 z-272m^2+192z^2+240z-33)}{(\rho_0+2\omega z)^2 G m^4}=\,0\,,
\eea
where we have used the equation of motion again.

For charged solutions, for example consider the M-charged solution at four derivative case, we find
\bea
&&\frac{\prd M}{\prd \rho_0}- T_H \frac{\prd S_{BH}}{\prd \rho_0}- \Omega_H \frac{\prd J}{\prd \rho_0}=-\frac{4 \rho_0(m^2+z m^2+\frac18)}{(\rho_0+2\omega z) G (8m^2+1)}=\,0\,\,\,,\nonumber\\
&&\frac{\prd M}{\prd \omega}- T_H \frac{\prd S_{BH}}{\prd \omega}- \Omega_H \frac{\prd J}{\prd \omega}=\frac{2 \rho_0 ((z-\frac32) m^2-\frac18)(m^2+z m^2+\frac18)}{(\rho_0+2\omega z) G m^4}=0.
\eea
In all cases in this paper, by using the equations of motion one can show that the equations (\ref{dform}) are correct.
\section*{Acknowledgment}
We would like to thank Mohammad Moghaddasi for discussion. We would like to thank Shahin Sheikh-Jabbari for his comments. This work was supported by Ferdowsi University of Mashhad under the grant 2/15734 (30/08/1389).

\begin{figure}[htp]
	\center
		\includegraphics[height=65mm]{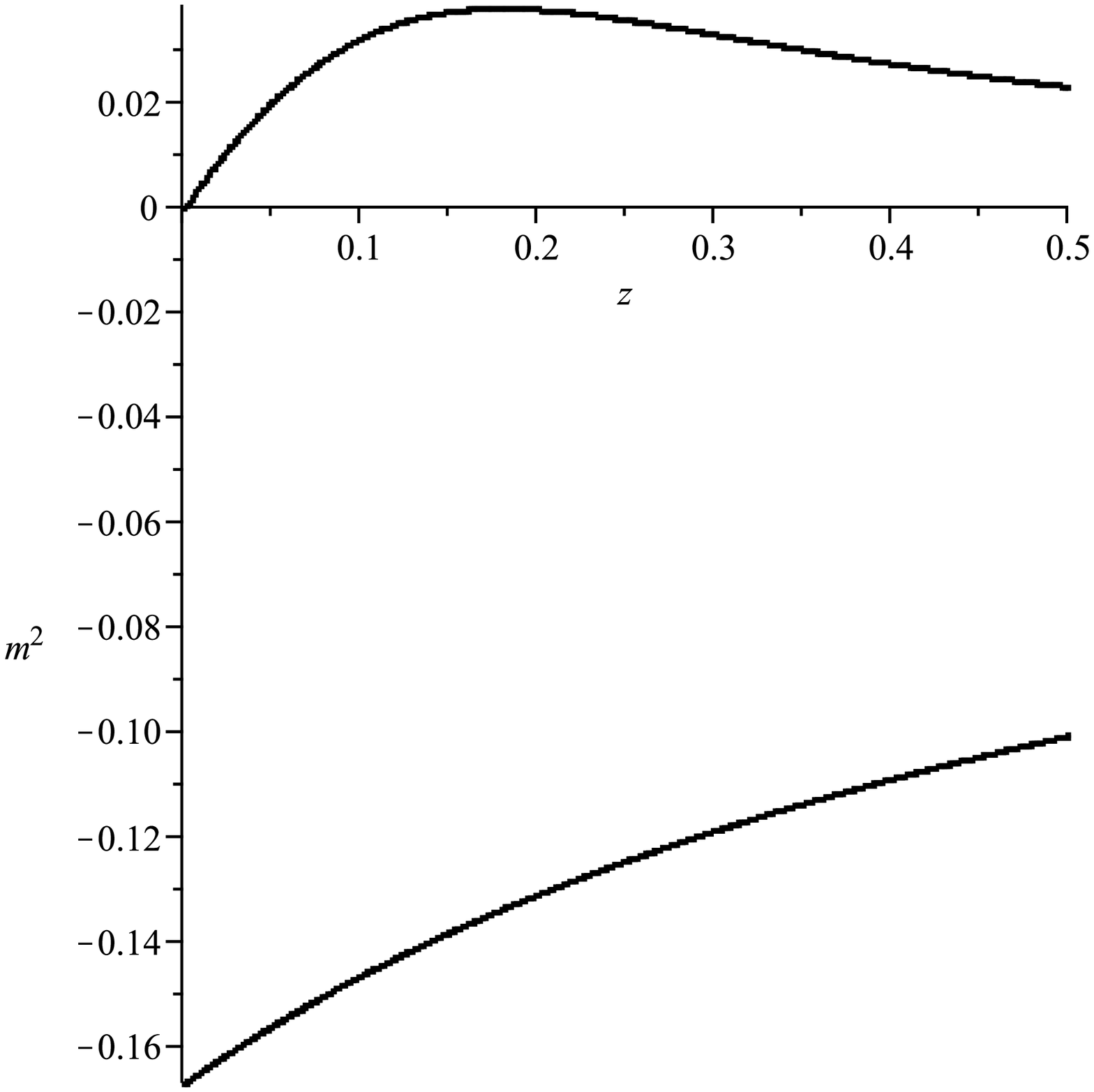}
		\includegraphics[height=65mm]{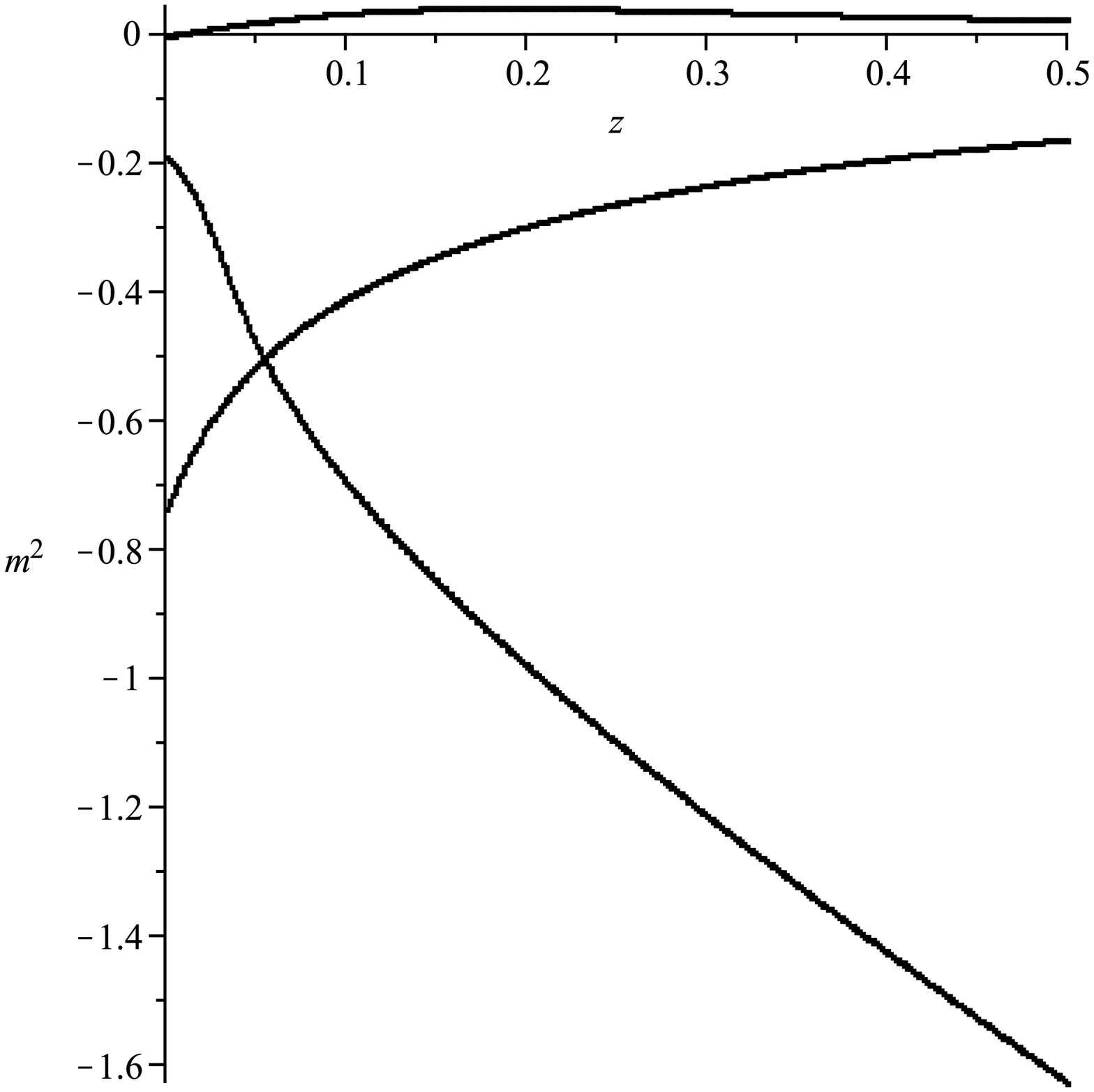}
	\caption{Left: Real roots of equation (\ref{poly1}). Right: Real roots of equation (\ref{poly2}).}
	\label{fig:fig1}
\end{figure}
\begin{figure}[htp]
	\center
		\includegraphics[height=65mm]{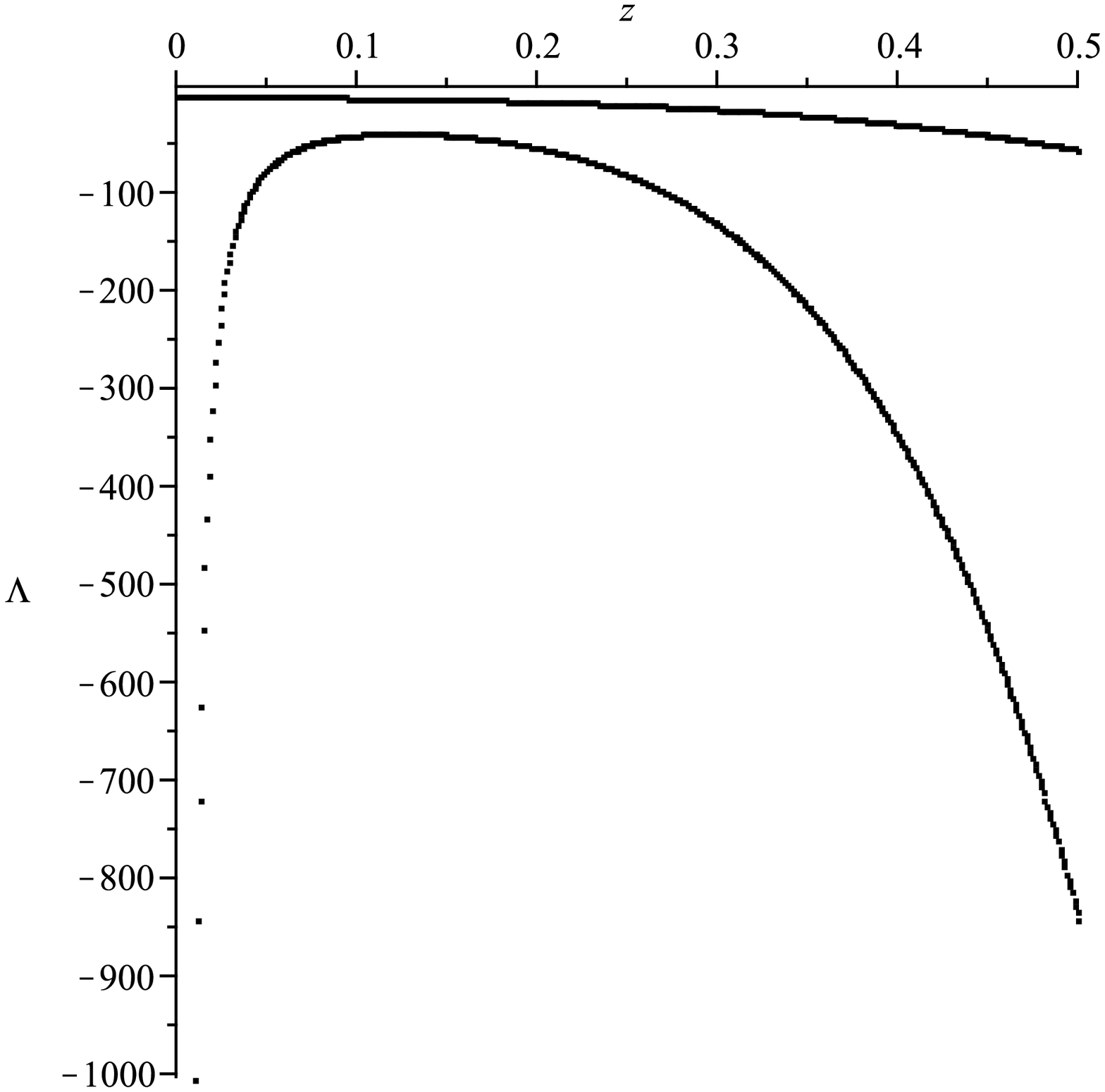}
		\includegraphics[height=65mm]{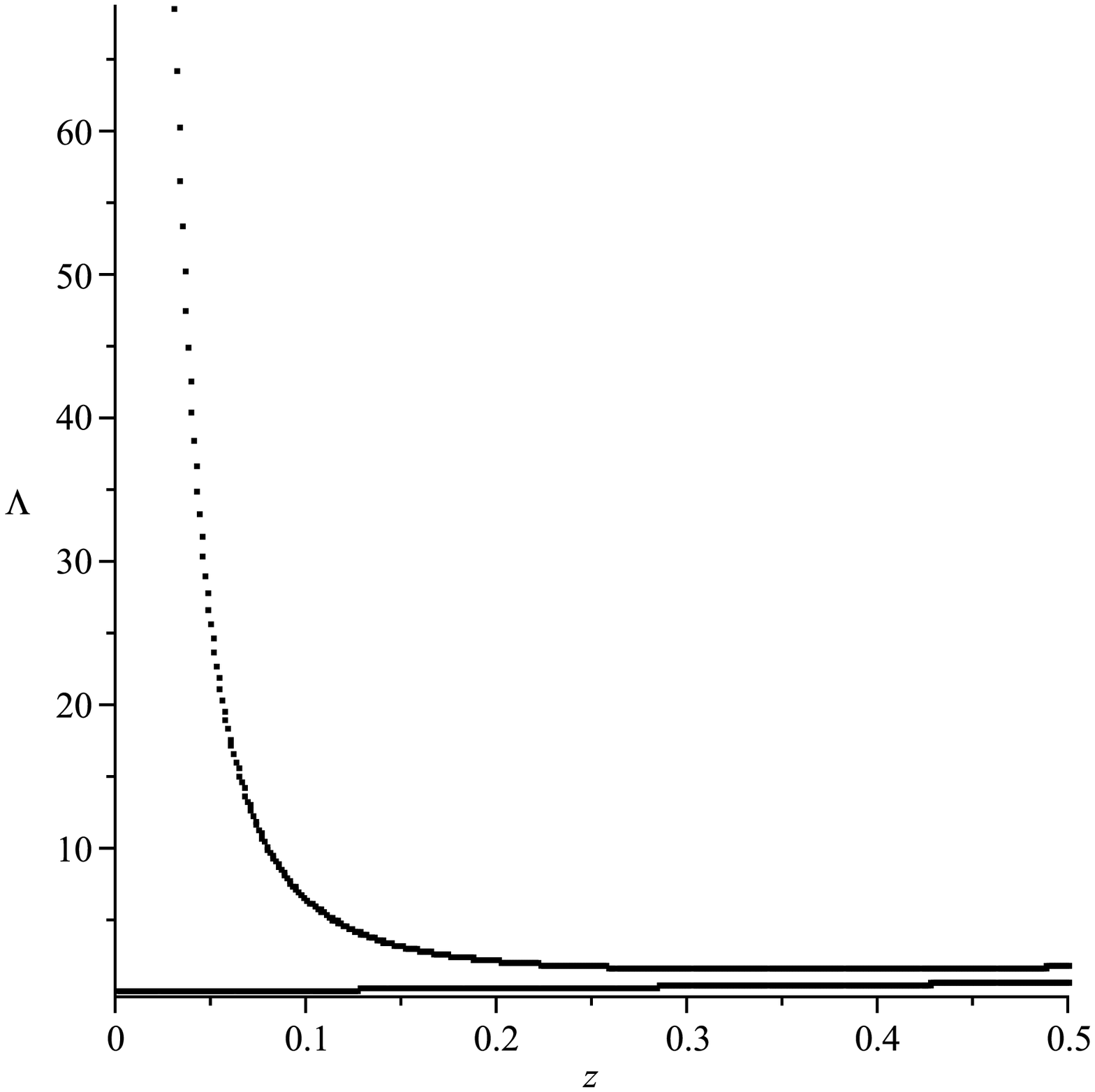}
	\caption{Cosmological constant for upper (left) and lower (right) signs of (\ref{Lambda3}).}
	\label{fig:fig2}
\end{figure}
\begin{figure}[htp]
	\center
	\includegraphics[height=65mm]{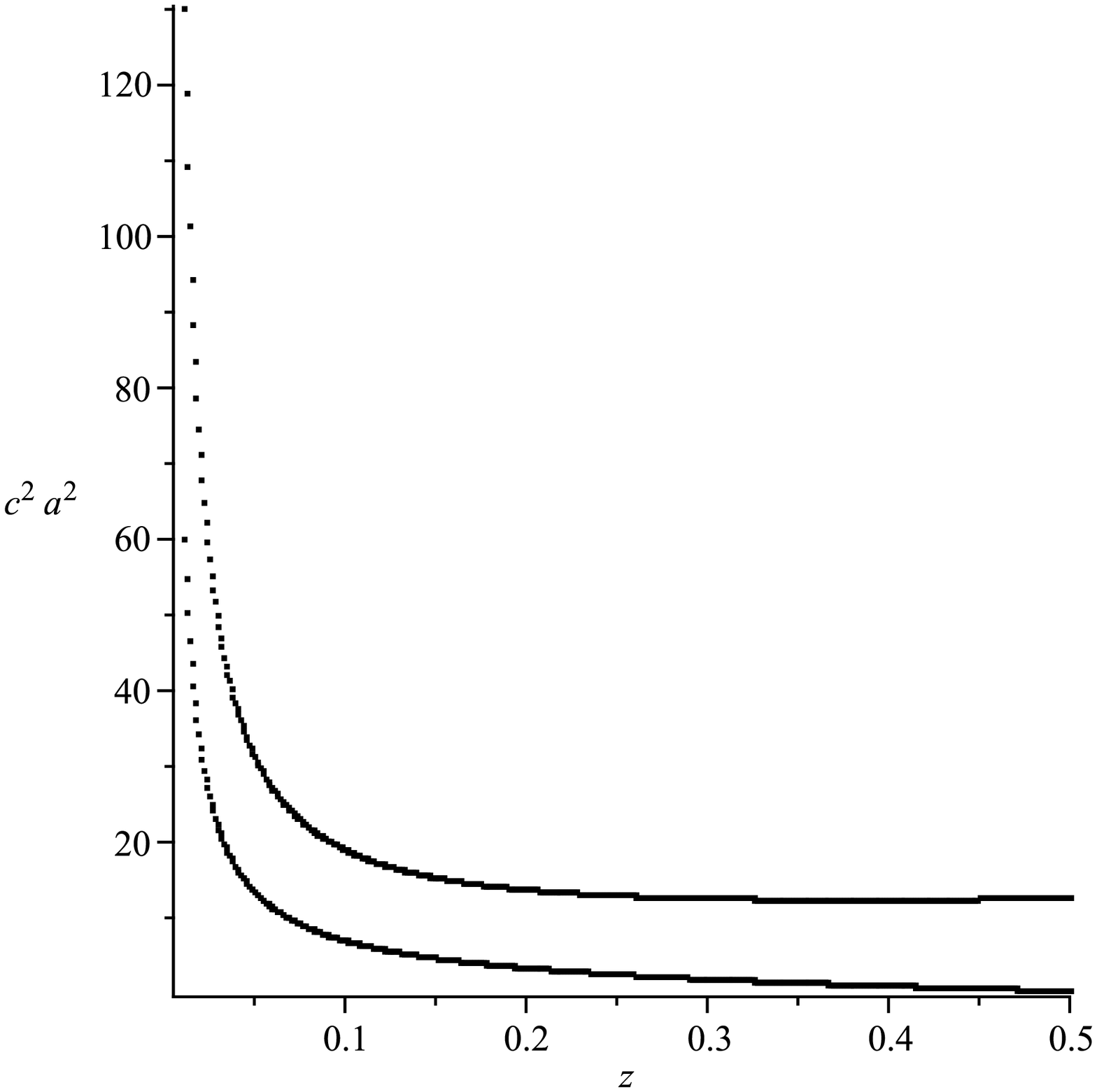}
	\includegraphics[height=65mm]{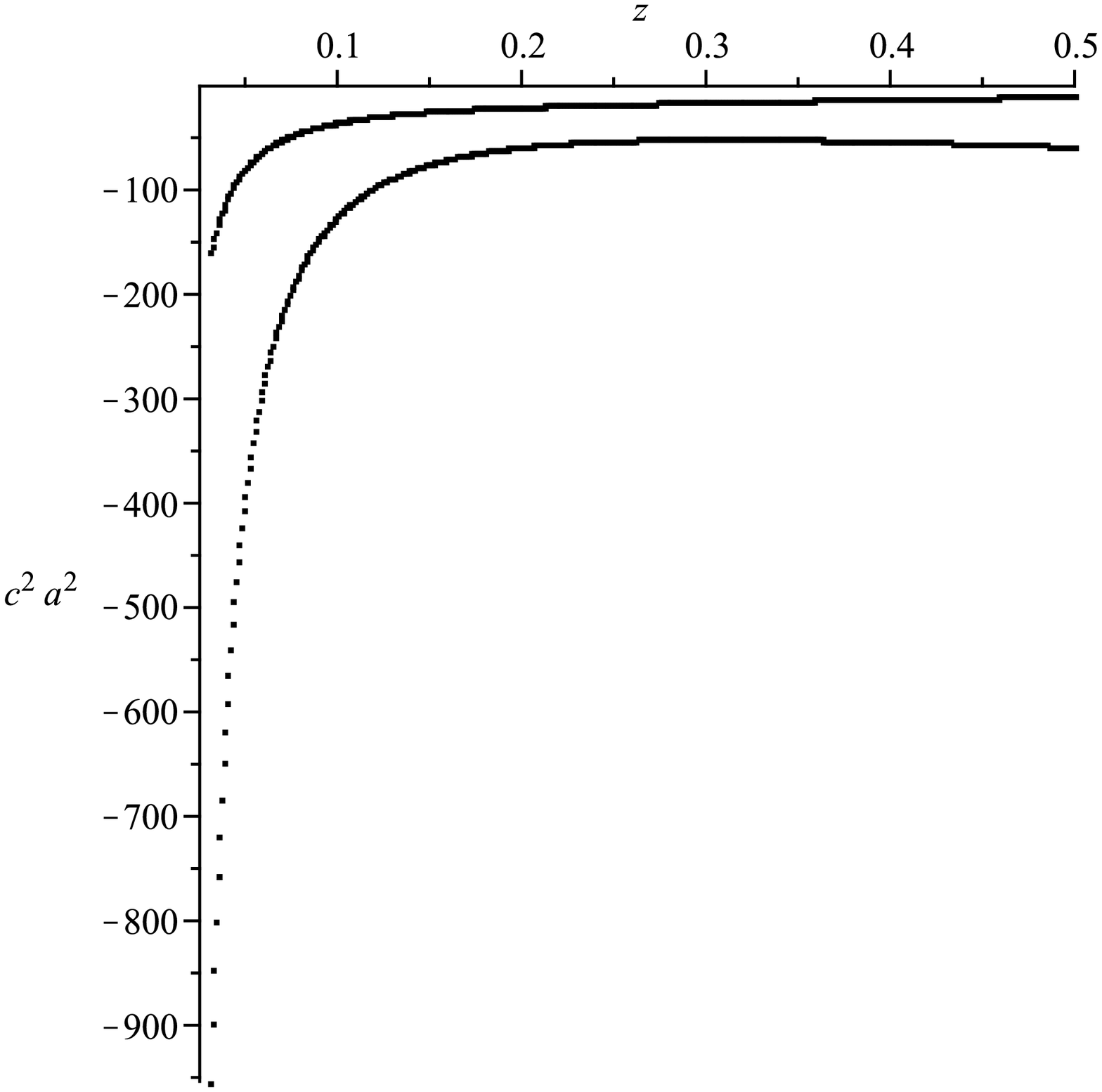}
	\caption{$a^2c^2$ in equation (\ref{c3}) for domain of $0<z<\frac12$.}
	\label{fig:fig3}
\end{figure}
\begin{figure}[htp]
	\center
	\includegraphics[height=65mm]{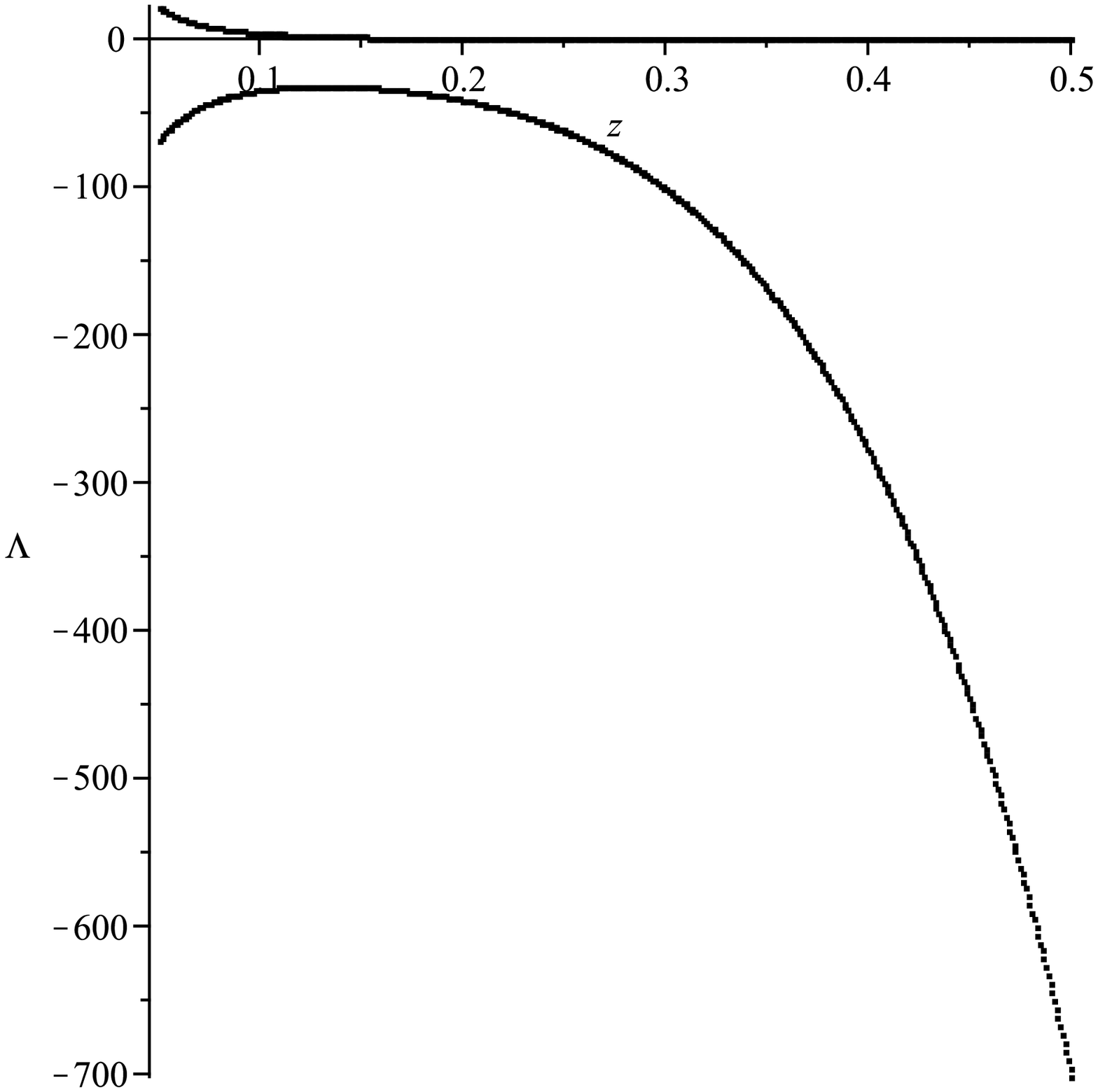}
	\caption{Values of cosmological constant in equation (\ref{Lambda4}) for domain of $0<z<\frac12$ and positive roots of (\ref{poly2}).}
	\label{fig:fig4}
\end{figure}
\begin{figure}[htp]
	\center
	\includegraphics[height=65mm]{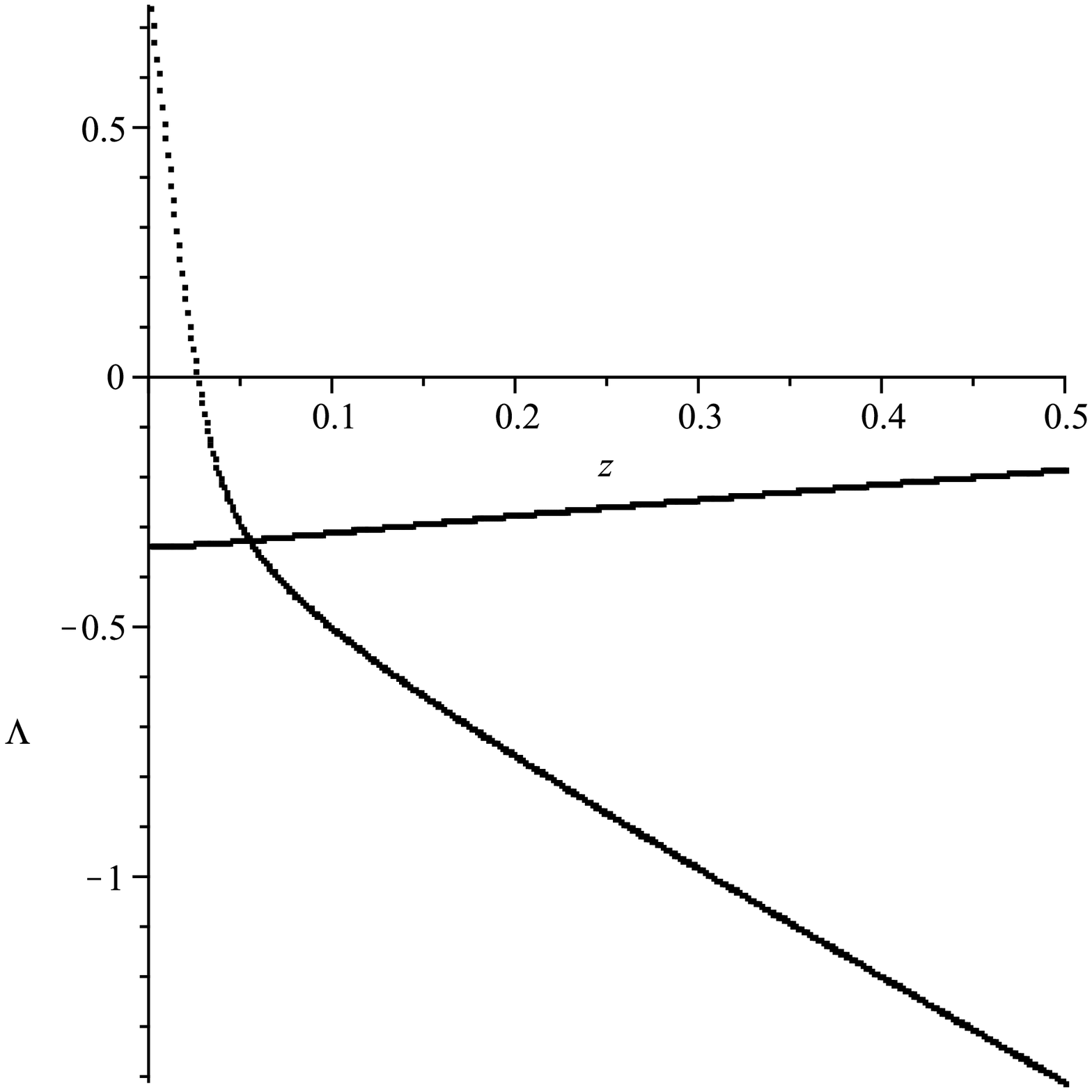}
	\includegraphics[height=65mm]{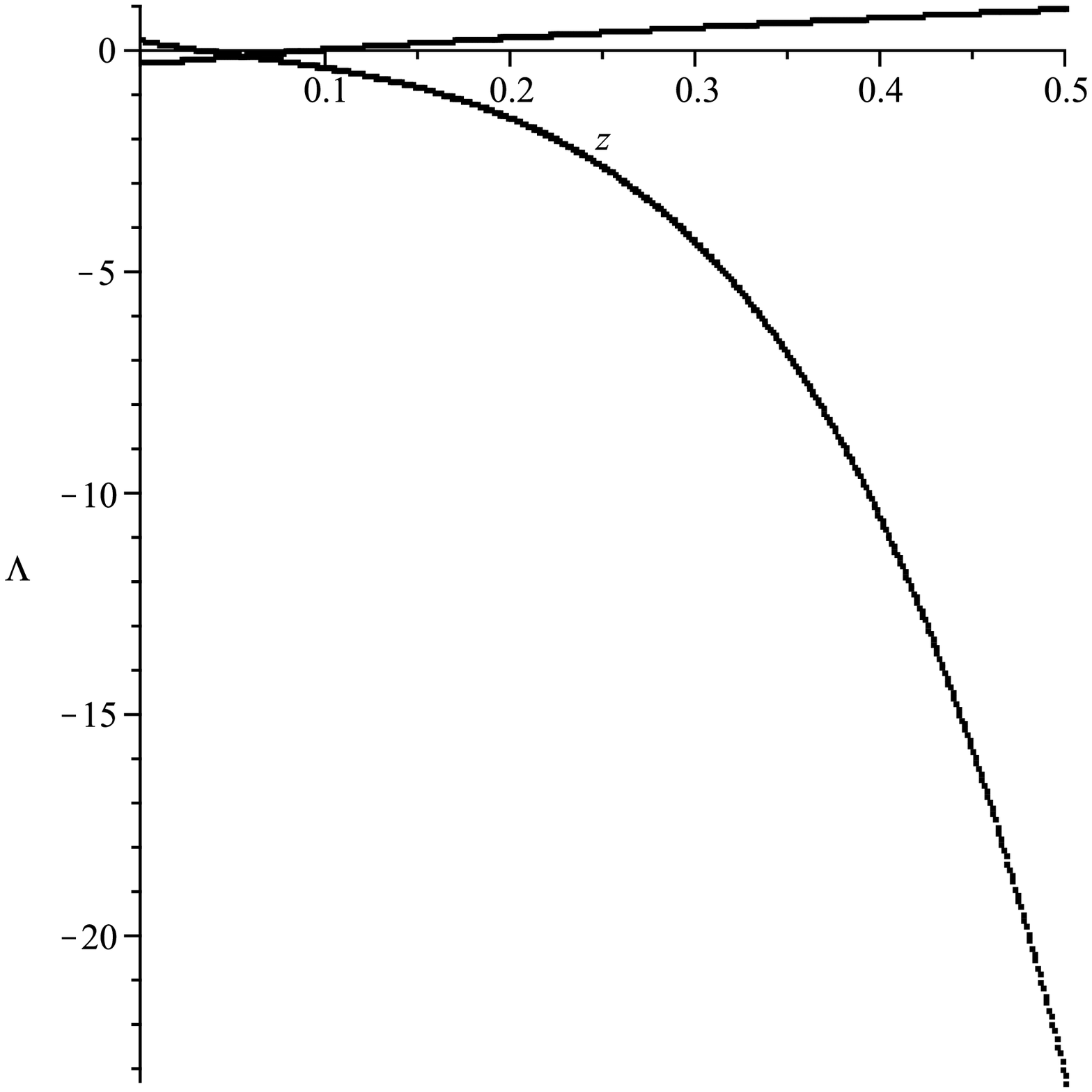}
	\caption{Values of cosmological constant in equation (\ref{Lambda4}) for domain of $0<z<\frac12$ and negative roots of (\ref{poly2}).}
	\label{fig:fig5}
\end{figure}
\end{document}